\documentclass[a4paper]{jpconf}
\usepackage{graphicx}
\usepackage{subfigure}
\begin{document}
\title{Structure and gap of low-$x$  (Ga$_{1-x}$In$_x$)$_2$O$_3$ alloys}

\author{M B Maccioni, F Ricci and V Fiorentini}

\address{CNR-IOM and Dept. of Physics, University of Cagliari, Cittadella Universitaria, 09042 Monserrato (CA), Italy }

\ead{mariabarbara.maccioni@dsf.unica.it, vincenzo.fiorentini@dsf.unica.it}

\begin{abstract}
We study  the electronic and local structural properties of pure and In-substituted $\beta$-Ga$_2$O$_3$  using density functional theory (DFT). Our main result is that  the structural energetics of In in Ga$_2$O$_3$ causes most sites to be essentially inaccessible to  In substitution, thus reducing the maximum In content in thi to somewhere between 12 and 25 \% in this phase. We also find that the gap variation with doping is essentially due to ``chemical pressure", i.e. volume variations with doping.
\end{abstract}

\section{Introduction}
Ga$_2$O$_3$ is   attracting interest recently as a material  for high-power transport and ultraviolet optical absorbers, owing to its wider band gap and larger electric breakdown voltage compared to e.g. GaN. Combined with In$_2$O$_3$ (already widely used as transparent conducting oxide), Ga$_2$O$_3$  originates a new materials system which is tunably insulating, easily $n$-doped (not so easily $p$-doped), and potentially magnetic (as In$_2$O$_3$ can be made ferromagnetic  \cite{alippi} through magnetic doping, the same may well apply to Ga$_2$O$_3$). 
Further,  the band-engineering and nanostructuration concepts from popular semiconductor systems such as, e.g., AlGaAs or InGaN  may be exported to these materials, and thus to a whole new region of high absorption energies and breakdown voltages.
 This may enable the design of devices based on Ga$_2$O$_3$/(Ga$_{1-x}$In$_x$)$_2$O$_3$ such as high-power field effect transistors and far-UV photodetectors or emitters.

\section{Ga$_2$O$_3$}
\begin{figure}
\centering
\includegraphics[width=9pc]{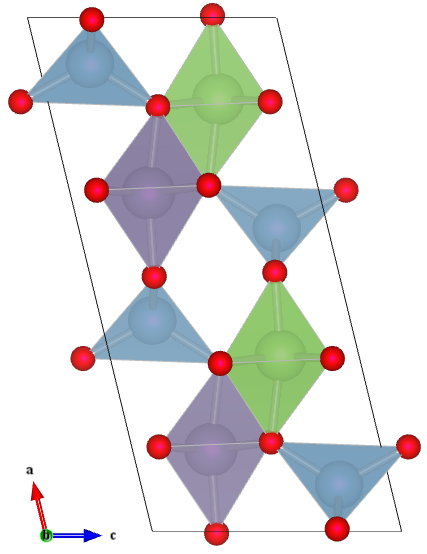}%
\qquad
\includegraphics[width=9pc]{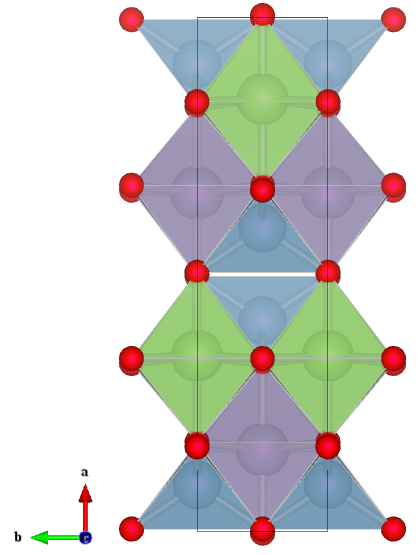}%
\qquad
\includegraphics[width=9pc]{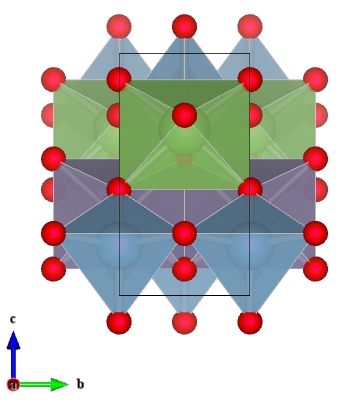}
\caption{\label{lattice}The structure of $\beta$-Ga$_2$O$_3$ can be seen as a collection of zigzag double chains of edge-sharing GaO$ _6 $ units (violet and green Ga-octahedra) linked by single chains of vertex-sharing GaO$ _4 $ (blue Ga-tetrahedra).}
\end{figure}

Gallium oxide, Ga$_2$O$_3$, exists in various polymorphs, the most stable being monoclinic $ \beta $-Ga$_2$O$_3$ at ambient condition  \cite{phases}.  The monoclinic phase, shown in Figure \ref{lattice}, belongs to the C2/m space group. The unit cell contains 20 atoms, with two crystallographically nonequivalent Ga atoms in tetrahedral and octahedral like coordinations in the lattice. Geometry and volume optimizations as well as electronic structure calculations have been done using  density functional theory (DFT) in the generalized gradient approximation (GGA), and the PAW method  as implemented in the VASP code  \cite{vasp}. The bulk Brillouin zone is sampled on a  4$\times$8$\times$6 Monkhorst-Pack grid.
The calculated lattice parameters  compare well with experiment  \cite{geller} (in parenthesis): $a$=12.46  \AA\, (12.23), $b$=3.08 \AA\, (3.04), $c$=5.88 \AA\, (5.80), $\theta$=103.65$^\circ $ (103.7).

The band structure of Ga$_2$O$_3$, similarly to other binary Ga compounds, has mainly O 2$p$ character in the upper valence band and Ga $s$ content in the bottom conduction band.  
The direct gap appears at the $\Gamma $ point. GGA underestimates it at about 2 eV, as expected of semilocal functionals. Adding an empirical self-energy correction \cite{fiore} involving the high-frequency dielectric constant, we obtain a gap of 4.2 eV, not far from the experimental range of 4.5-5 eV. 

Surprisingly, the precise value of the gap is still uncertain even in recent work  \cite{zang}. The reason for this is probably the significant anisotropy of the absorption, which we have analyzed, and will report elsewhere  \cite{ricci},  with hybrid functionals and pseudo self-interaction corrections (known to be free of the typical LDA/GGA gap errors). For the present purposes, we just note that these advanced methods confirm that a  direct minimum gap at zone center between 4.2 eV (hybrids) and 4.7 eV (self-interaction-correction), and also confirm the pressure derivative of the gap to be 3 meV/kbar essentially as in GGA (see below). 

\section{Low-In-content alloying}
Because unalloyed In and Ga oxides have different structures (bixbyite and monoclinic $\beta$, respectively) the high-In and low-In-content alloying limits will behave quite differently, and at intermediate concentrations the two phases are likely to  mix in an complicated way. The experimental alloying of Ga$_2$O$_3$ with In$_2$O$_3$ indeed faces significant limitations  \cite{zang}, with  $\beta$-Ga$_2$O$_3$-like and bixbyite-like X-ray spectra at low $x$ and high $x$ respectively, and a mixed-phase region at midrange $x$. In particular the $\beta$-Ga$_2$O$_3$-like phase persists only up to about 15\% or so  \cite{zang}. Thus, keeping in mind that the large-$x$ end of the alloying spectrum will have to be treated differently, here we  tackle the low-$x$ end  substituting In for Ga in $\beta$-Ga$_2$O$_3$ at nominal concentrations of 3, 6, 9, and 12 \% (one to four In atoms per  80-atom or 32-cation supercell). Our results naturally suggest an interpretation for the observed behavior.  

We optimize (in volume, shape, and internal coordinates) supercells of Ga$_2$O$_3$  with  In$\rightarrow$Ga substitutions  sampling some of the various possible octahedral and tetrahedral sites and combinations thereof as function of composition (i.e. In concentration);  we use a 2$\times$4$\times$2 k-point grid. The calculations at 3\% In (one "isolated" In atom per 80-atom cell) show that  In only substitutes octahedral Ga: tetrahedral sites are ruled out by a huge excess energy cost of 1.1 eV. Therefore, half the cation sites are essentially inaccessible to In, and hence the amount of In that can  actually be incorporated into Ga oxide is automatically halved -- to put it differently, all available sites would be occupied already at 50\% nominal In content. 
\begin{figure}[h]
\begin{center}
\includegraphics[width=14pc]{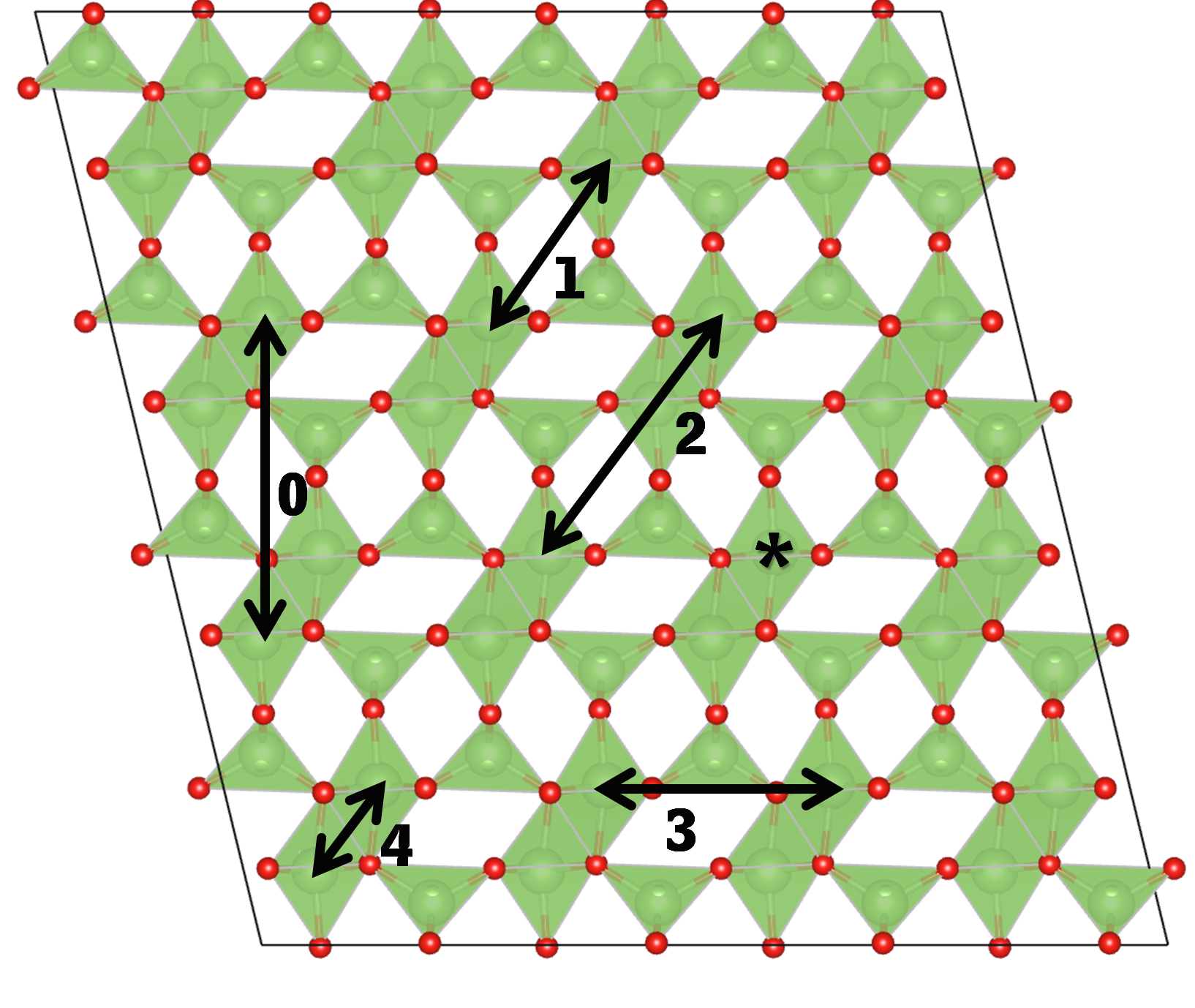}
\includegraphics[width=17pc]{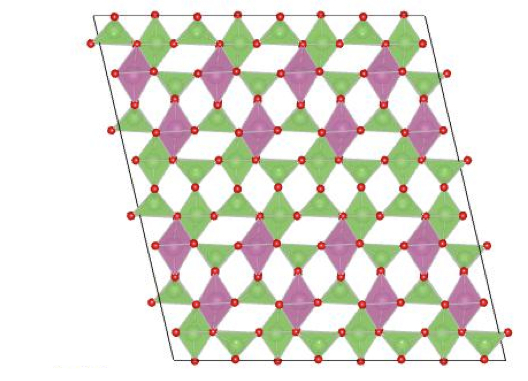}
\end{center}
\caption{\label{gapall}Left: different pairings patterns of In in Ga$_2$O$_3$. Right: structure of the 12\% sample.}
\end{figure}

Even reaching this limit, though, is quite unlikely.
Indeed, In incorporation is not arbitrary in terms of configuration. At 6\% In content, i.e  two In atoms per supercell, one can estimate the energetics of pairing (or, rather, non-pairing) of In in  Ga$_2$O$_3$. 
In Figure \ref{gapall}, left, we display the preferential couplings. The energy of configuration '0' is chosen as zero; the structures numbered '1' to '4' are in progressively unfavorable energetic order, with   '1' at 16 meV, '2' at 50 meV, '3' at 100 meV, '4' at 125 meV. In the configuration labeled '*', In atoms occupy adjacent octahedra; this structure is 250 meV higher than the reference. 

Clearly, 
In atoms  tend to avoid one another, and it is likely that at the common growth temperatures  of 850 K the typical configurations will be such as our  '0' and  '1' above. Inspecting the structure, this suggests that well below a half, and probably closer to a quarter, of the octahedral sites can be occupied by In with a reasonable  energy cost; when these are filled, the  formation of some mixed $\beta$/bixbyite phase may be preferable to substitution in the $\beta$ phase. This brings the effective solubility in the original $\beta$-Ga$_2$O$_3$ structure down to  between 12\% and 25\%  as found in experiment  \cite{zang}.

Consistently with  the above configurational restrictions on pairing, the  admissible arrangements at 9\% and especially 	12\% In content  are few. For example, the structure used at 12\% is in Figure \ref{gapall}, right.
The resulting structures are probably a fairly decent model of the alloy, given the very limited configurational freedom of In already at these concentrations. (Cluster-expansion work is ongoing on miscibility at finite temperature and will be reported elsewhere.) 

\begin{figure}[h]
\begin{center}
\includegraphics[width=18pc]{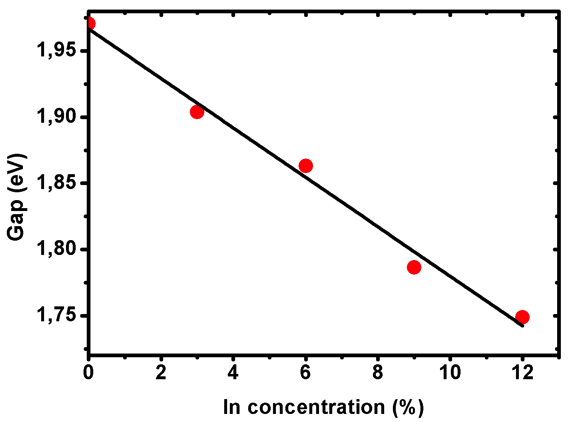}
\includegraphics[width=18pc]{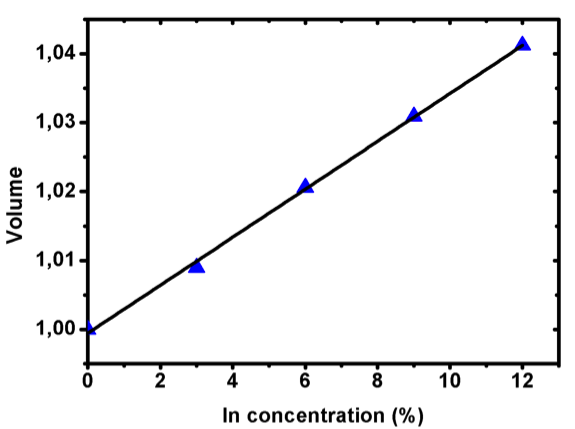}
\caption{\label{gapall1} Left: direct gap at zone center as function of In concentration. Right:
normalized volume of In-doped  Ga$_2$O$_3$ as function of concentration.}
\end{center}
\end{figure}

We calculated the gap and the volume change with concentration of (Ga$_{1-x} $In$ _{x} $)$_2$O$_3$ in the 80-atom supercell for the structures mentioned. The gap is in Figure \ref{gapall1}, left, and the volume is in the same Figure, right. Although the gap is underestimated (a token of using GGA), its concentration change of 17 meV/$\% $ is quite similar to 20 meV/\% experimentally \cite{zang}.
Using the the volume change with $x$ and the bulk modulus, we can evaluate the concentration derivative as a  pressure derivative, obtaining 2.5 meV/kbar. This is similar to the Ga$_2$O$_3$ value of  3 meV/kbar  \cite{ricci,Blanco}, which suggests that the gap is mainly affected by volume change, and marginally by other factors. In this light, the agreement with experiment therefore falls in line with expectations from previous work \cite{vf92}.

\section{Conclusions}
In summary, we have performed first-principles calculations on  the electronic and local structural properties of In-containing Ga$_2$O$_3$. The energetics of In in Ga$_2$O$_3$ limits the maximum In content to somewhere between 12 and 25 \%. We also find that the gap variation with doping is essentially due to ``chemical pressure", i.e. volume variations with doping.

\ack
M.B.M. gratefully acknowledges Sardinia Regional Government for the financial support of her PhD scholarship (P.O.R. Sardegna F.S.E. Operational Programme of the Autonomous Region of Sardinia, European Social Fund 2007-2013 -
Axis IV Human Resources, Objective l.3, Line of Activity l.3.1). This work was also supported in part by MIUR-PRIN 2010 project {\it Oxide}, Fondazione Banco di Sardegna and CINECA grants.


\section*{References}
\begin{thebibliography}{9}
\bibitem{alippi} Alippi P, Cesaria M, and Fiorentini V 2014 \textit{Phys. Rev. B} \textbf{89} 134423. 
\bibitem{phases} Edwards D F 1998 \textit{Handbook of Optical Constants of
Solids} Academic Press, \textbf{III} 753
\bibitem{geller} Geller S 1960 \textit{J. Chem. Phys.} \textbf{33} 676
\bibitem{vasp} Kresse G and Furthmuller J 1996 \textit{Phys. Rev. B}
\textbf{54} 11169
\bibitem{fiore} Fiorentini V and Baldereschi A 1992 \textit{J. Phys.: Condens. Matter} \textbf{4} 5967
\bibitem{zang} Zhang F et al 2014 \textit{Solid State Comm.} \textbf{186} 28
\bibitem{ricci} Ricci F et al 2014, to be published.
\bibitem{Blanco} He H et al 2006 \textit{Phys. Rev. B} \textbf{74} 195123. 
\bibitem{vf92} Fiorentini V 1992 \textit{Phys. Rev. B} \textbf{46} 2086



\end{thebibliography}
\end{document}